\documentclass[twocolumn,showpacs,preprintnumbers,pra,floatfix]{revtex4-2}
\usepackage{graphicx}% includes figure files
\usepackage{epsfig}% include figure files
\usepackage{bm}% bold math
\usepackage{amssymb,amsmath}
\usepackage{lipsum}
\usepackage{multirow}
\usepackage{nth}

\def\4he{$^4$He}
\def\3he{$^3$He}
\def\cm3{cm$^{-3}$}

\begin{document}

\title{Part-per-billion measurement of the  $4^2S_{1/2} \rightarrow 3^2D_{5/2}$ electric quadrupole transition isotope shifts between $^{42,44,48}$Ca$^+$ and $^{40}$Ca$^+$}

\author{Felix W. Knollmann$^1$, Ashay N. Patel$^{1,2}$, and S. Charles Doret$^1$\footnote{scd2@williams.edu}}
\affiliation{
$^1$Williams College Dept. of Physics, Williamstown, MA  01267\\
$^2$Caltech Division of Physics, Mathematics, and Astronomy, Pasadena, CA  91125 }
% include data for PRL submission
\date{June 10, 2019}

\begin{abstract}
\noindent We report a precise measurement of the isotope shifts in the $4^2$S$_{1/2} \rightarrow 3^2$D$_{5/2}$ electric quadrupole transition at 729~nm in $^{40 - 42,44,48}$Ca$^+$.  The measurement has been made via high-resolution laser spectroscopy of co-trapped ions, finding measured shifts of 2,771,872,467.6(7.6), 5,340,887,394.6(7.8), and 9,990,382,525.0(4.9) Hz  between $^{42,44,48}$Ca$^+$and $^{40}$Ca$^+$, respectively.  By exciting the two isotopes simultaneously, using frequency sidebands derived from a single laser, systematic uncertainties resulting from laser frequency drifts are eliminated.  This permits far greater precision than similar previously published measurements in other alkaline-earth systems.  The resulting measurement accuracy provides a benchmark for tests of theoretical isotope shift calculations, and also offers a step towards probing new physics via isotope shift spectroscopy.  
\end{abstract}

% these may not be accurate
% include these for PRL submission

\maketitle
\section{Introduction}

\noindent Isotope shift spectroscopy has long contributed to our understanding of atomic and nuclear structure \cite{Breit58, King84, CMP16}.  Such shifts result from small perturbations to electronic transition energies due to the variation in mass and size/shape of the nuclei of different isotopes.   Precise isotope shift measurements thus provide sensitive tests of both relativistic many-body physics used to calculate electron wavefunctions and ever-improving theory of nuclear structure \cite{BDN13}.  Aside from support for theory, measurements of isotope shifts serve as the principal probe of mean-square nuclear charge radii \cite{Sta66, BC95, KN03}, help to tease out information about nuclear ground-state properties encoded in atomic spectra \cite{CMP16}, contribute to our understanding of a variety of important parameters in the Standard Model \cite{DJS05, KBG18e, MVB11e, DOP17e}, and are of critical importance in understanding variations in isotopic abundances in astrophysical sources applicable to searches for spatial and temporal variation of fundamental constants \cite{KKB04e, BDF11e, MB14, KK07}.

Typically an isotope shift is modeled as arising from two different contributions.  The ``mass shift'' stems from the differing nuclear masses of various isotopes and is generally proportional to the relative mass change between isotopes, while the ``field shift'' is due to variation in the spatial distribution of nuclear charge.  It has long been known  that, if one assumes that the field shift is proportional to the difference in the mean-squared nuclear charge distribution between isotopes,  the isotope shifts in two sets of transitions in the same isotopes can be linearly related to one another in a ``King plot'' \cite{King63}.  With the notable exception of near-degenerate levels in samarium \cite{GIN81e}, virtually all existing isotope shift measurements between spin-zero isotopes are consistent with this linear relationship, known as ``King's linearity.''  However, recent work \cite{DOP17e, FFP17e, MTY17, BBD18e,  FGV18} has forecast a possible breaking of King's linearity due to new physics beyond the Standard Model, for example spin-independent electron-neutron couplings arising from conjectured new lightweight bosons.  Such non-linearities may also arise from higher-order contributions within the Standard Model, such as nuclear polarizabilities \cite{FGV18, PS82, Seltzer69, BBP87e,TFR85}.  It is thus highly desirable to obtain new, highly precise (Hz-level) measurements of isotope shifts in a variety of systems to enable discrimination between new physics and Standard Model contributions \cite{BBD18e, FGV18}. 

\begin{figure}[h!]
\includegraphics[width=7.0 cm]{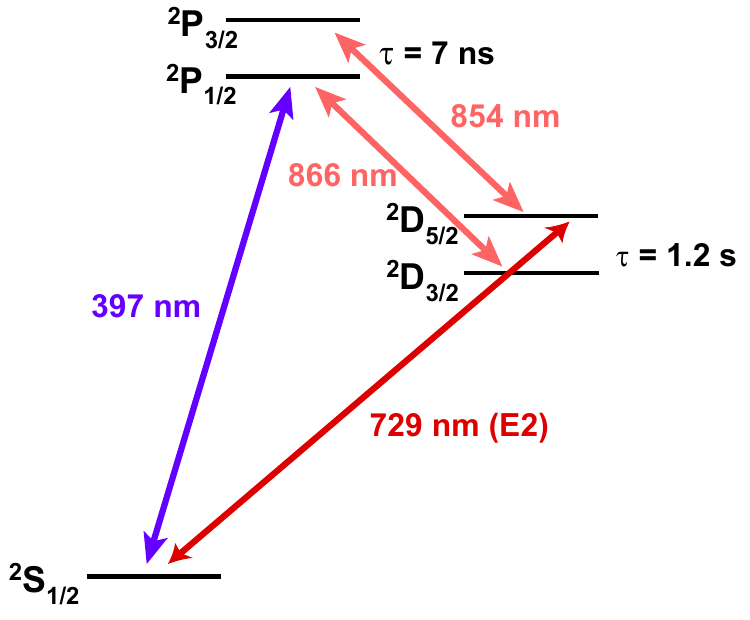} % Making this figure any bigger pushes the next figure off the following page. 
\caption{\label{fig:levels} (color online) Level diagram for nuclear spin-zero isotopes of Ca$^+$, with natural lifetimes listed.  The 397~nm transition is used for Doppler cooling and fluorescence detection, while metastable $^2$D$_{3/2, 5/2}$ levels are repumped by transitions at 866~nm and 854~nm, respectively.}
\end{figure}

In principle any pair of transitions could be used in searching for a King nonlinearity, but dipole-forbidden transitions in trapped ions represent a particularly appealing candidate system.  Calcium is ideally suited to such a search; it features a long chain of stable isotopes ($^{40,42,43,44,46,48}$Ca),  five of which are nuclear spin-zero and thus lack hyperfine interactions which are known sources for King nonlinearities within the Standard Model \cite{PS82}.    A suite of dipole-allowed transitions (Fig.~\ref{fig:levels})  permit straightforward laser-cooling and trapping as well as application of 
 \begin{figure*} %* makes this two-column.  This is way out of whack with where I want it to appear in the text, but this is where the code has to be located to get it to place properly.
\includegraphics[width=.95\textwidth]{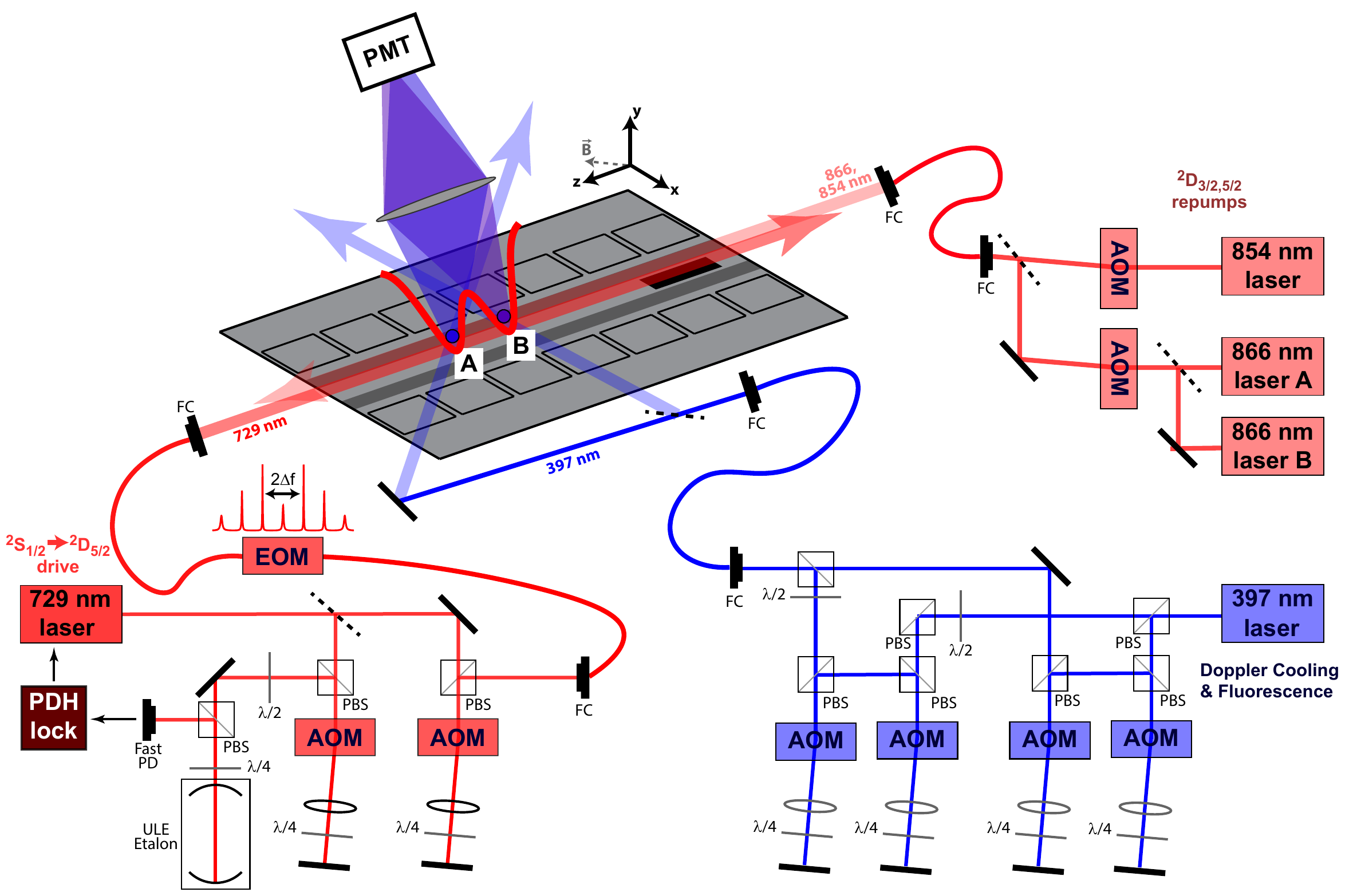}
\caption{\label{fig:apparatus} (color online) Simplified schematic of the experimental apparatus.  Two Ca$^+$ isotopes (here labeled ``A'' and ``B'') are co-trapped in 100~$\mu$m-separated potential wells and simultaneously driven by an ultrastable ECDL at 729~nm.  A fiber-EOM generates frequency sidebands which span the transition isotope shift, while an AOM allows the laser to be scanned across the $4^2$S$_{1/2} \rightarrow 3^2$D$_{5/2}$ transition.  A weak magnetic field in the plane of the trap sets the ions' quantization axis.  Standard laser setups at 397, 866, and 854~nm are used for Doppler cooling, fluorescence detection, and repumping/clean-out of the $3^2$D$_{3/2,5/2}$ states.  Not pictured are lasers and associated with photoionization for trap loading (423 and 375~nm) and frequency stabilization of lasers at 397, 423, and 866~nm (transfer cavity lock to a commercial stabilized HeNe \cite{LSD91,Zhao98}).}
\end{figure*}
techniques from the arena of quantum information processing, enabling spectroscopic measurements of exquisite precision \cite{CBK09e, WGW14e, GWW15e}.  Isotope shift measurements in the calcium system have already borne fruit in several areas, including improved understanding of stellar spectra \cite{NBI98e, CH04, CH05}, determination of nuclear charge radii \cite{PBB84e, MYW92e, VSL92e, GBB16}, and trace-isotope analysis \cite{LW03}.  Of particular interest here are the two long-lived metastable D$_{5/2,3/2}$ states \cite{SID06e}  which give rise to narrow electric quadrupole transitions at 729 and 732~nm, respectively.  A King plot comparing isotope shifts in these transitions, or alternatively comparing either one to the $^1$S$_0 \rightarrow ^3$P$_1$ intercombination line in neutral calcium \cite{DSL05e}, would yield a sensitive probe of King's linearity.  

Here we present measurements of isotope shifts in the $4^2$S$_{1/2} \rightarrow 3^2$D$_{5/2}$ 729~nm electric quadrupole transition in $^{40 - 42,44,48}$Ca$^+$.  In contrast to similar measurements in other alkaline-earth systems which trapped and probed one ion at a time \cite{ZYD95e, BGG03e, LBC11}, we co-trap the two isotopes and simultaneously interrogate them with frequency sidebands derived from a single laser \cite{Srmeas}.  This method eliminates systematic uncertainties due to laser frequency instability, allowing the present measurement to improve on the precision of these analogous measurements by more than two orders of magnitude.  Our measurement thus serves as a stringent test for atomic and nuclear structure calculations in alkali-like systems, and contributes to an exacting test of King's linearity in the calcium system (see Fig.~\ref{fig:KP} below).

\section{Experimental Methods}

We co-trap two isotopes of calcium in a surface-electrode RF-Paul trap  \cite{LRH04e} in an apparatus (Fig.~\ref{fig:apparatus}) similar to those described elsewhere \cite{DAW12e, HRB08}.  In brief, home-built external cavity diode lasers (ECDLs) at 423 and $\sim$375~nm allow ions to be loaded via species-selective photoionization from a natural-abundance calcium vapor produced by a resistively heated oven.  Once trapped the ions are shuttled  to one of two storage wells located roughly 1.5~mm away from the trap's loading slot.  After both isotopes have been successfully loaded the two storage wells are partially merged into a double-well with minima spaced by approximately 100~$\mu$m and characteristic axial secular frequencies of 1.4~MHz.  A DC quadrupole potential rotates the trap axes to permit 3D Doppler cooling and splits the radial center-of-mass modes to secular frequencies of 2.5 and 3.5~MHz.  Additional home-built ECDLs at 397~nm and 866~nm enable Doppler cooling and fluorescence detection.  

To permit simultaneous Doppler cooling and fluorescence detection on two isotopes we span the isotope shift on the 397~nm transition with a sequence of double-passed $\sim$200~MHz AOMs.  Two separate lasers at 866~nm span the larger isotope shift on the $3^2$D$_{3/2} \rightarrow 4^2$P$_{1/2}$ repumping transition.  Rabi flopping on the 729~nm  $4^2$S$_{1/2} \rightarrow 3^2$D$_{5/2}$ transition indicates that both ions are cooled to $2(1) \times T_{\textrm{Doppler}}$, regardless of isotopic pairing.   All lasers are piped to the ion trap via polarization-maintaining single-mode optical fibers, with translation-stage-mounted-singlet lenses used to focus light from each fiber onto the trapped ions.  The 397 nm beam is split in two prior to being focused into the vacuum chamber.  One beam, broadened by a cylindrical lens, continuously illuminates both minima of the double-well potential, while a second beam can be steered to any location in the trap via a fast steering mirror.  The 866~nm repumping light is directed along the trap's axial direction, illuminating all possible trapping locations equally.  

\begin{figure}[h!]
%May want to make this figure two-column
\includegraphics[width=7.7 cm]{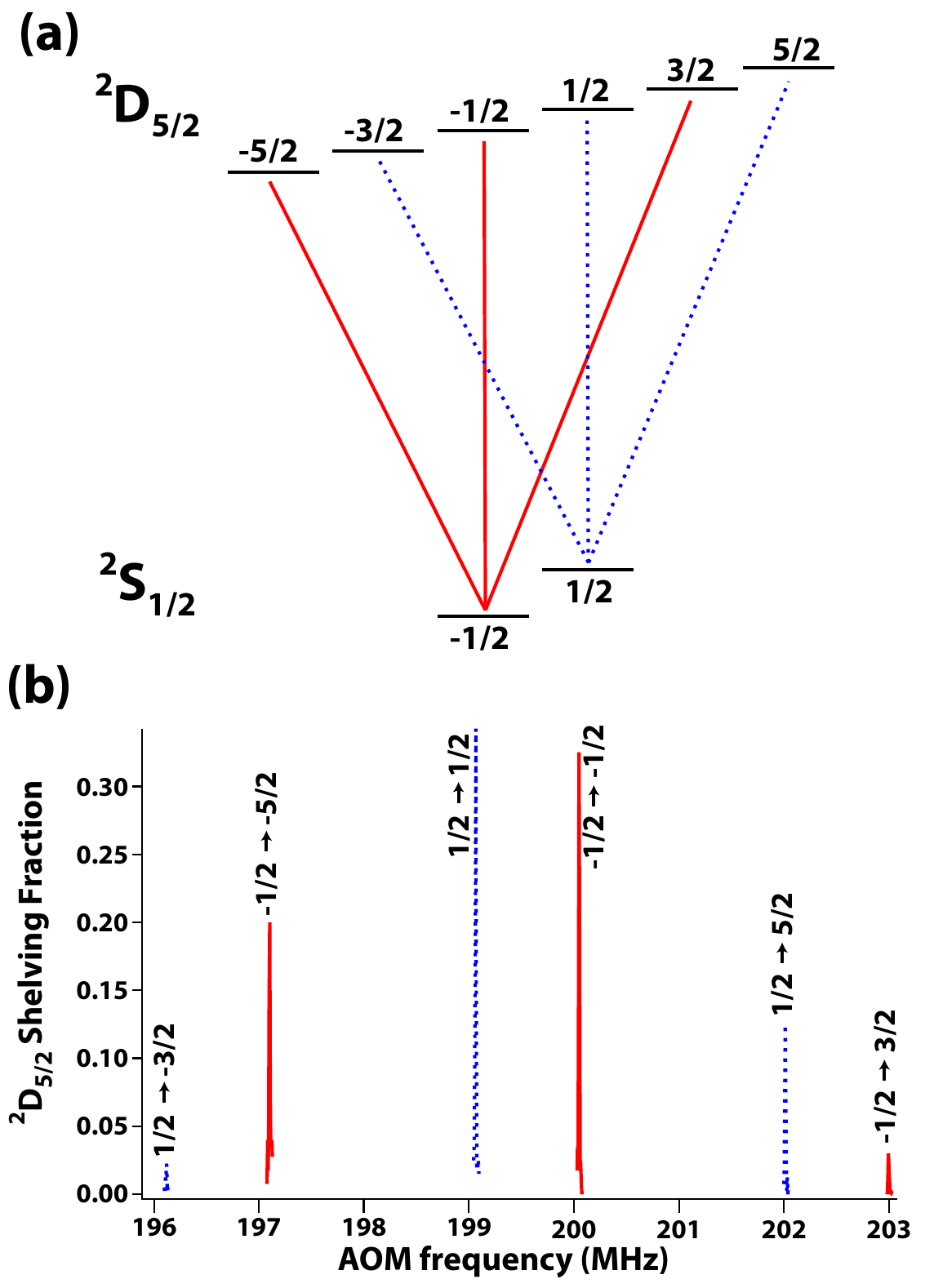}
\caption{\label{fig:transitions} (color online) Spectra at 729~nm.  (a,b) Level structure and allowed transitions for the $4^2$S$_{1/2} \rightarrow 3^2$D$_{5/2}$ transition with quantization axis and 729~nm laser propagation as described in Fig.~\ref{fig:apparatus}.  Isotope shift measurements are made on the strongest lines: $|^2\textrm{S}_{1/2}, m_J=\frac{1}{2}\rangle \rightarrow |^2\textrm{D}_{5/2},  m_J=\frac{1}{2}\rangle$ and  $|^2\textrm{S}_{1/2},  m_J=-\frac{1}{2}\rangle \rightarrow |^2\textrm{D}_{5/2},  m_J=-\frac{1}{2}\rangle$. }
\end{figure}

Ions are probed on the $4^2$S$_{1/2} \rightarrow 3^2$D$_{5/2}$ electric quadrupole transition at 729~nm using a commercial ECDL (Toptica DL100) stabilized using a Pound-Drever-Hall lock \cite{DHK83e} to an in-vacuum, temperature stabilized, ultra-low-expansion etalon.  A double-passed $\sim$80~MHz AOM between the laser and the etalon is slowly tuned to remove slow linear drifts that result from aging of the etalon, leaving the laser with a residual frequency instability of $<10$~kHz/day.  A double-passed $\sim$200~MHz AOM is used for switching and frequency control of the beam sent to the ions, while a DC-15~GHz broadband fiber EOM is used to create frequency sidebands which span the $4^2$S$_{1/2} \rightarrow 3^2$D$_{5/2}$ transition isotope shift.  After reaching the trap via the EOM's fiber pigtail, the 729~nm light is aligned antiparallel to the 866~nm beam so as to equally illuminate both minima of the double-well trapping potential.  A magnetic field of 1.755(1)~G oriented at 45$^\circ$ to the 729~nm laser's propagation direction sets the ions' quantization axis.  In combination with this geometry, selecting a 729~nm laser polarization perpendicular to the quantization field suppresses $\Delta m = \pm1$ transitions, leaving the six transitions shown in Fig.~\ref{fig:transitions}.  This geometry has the additional benefit of maximizing the strength of the $\Delta m = 0$ transitions which have the smallest magnetic field sensitivity.   A single additional home-built ECDL at 854~nm, centered at a frequency equidistant from the relevant transitions in both isotopes and co-propagating with the 866~nm lasers, is used to quench unwanted population in the $3^2$D$_{5/2}$ state.

\begin{figure}[t!]
%May want to make this figure two-column
\includegraphics[width=8.5 cm]{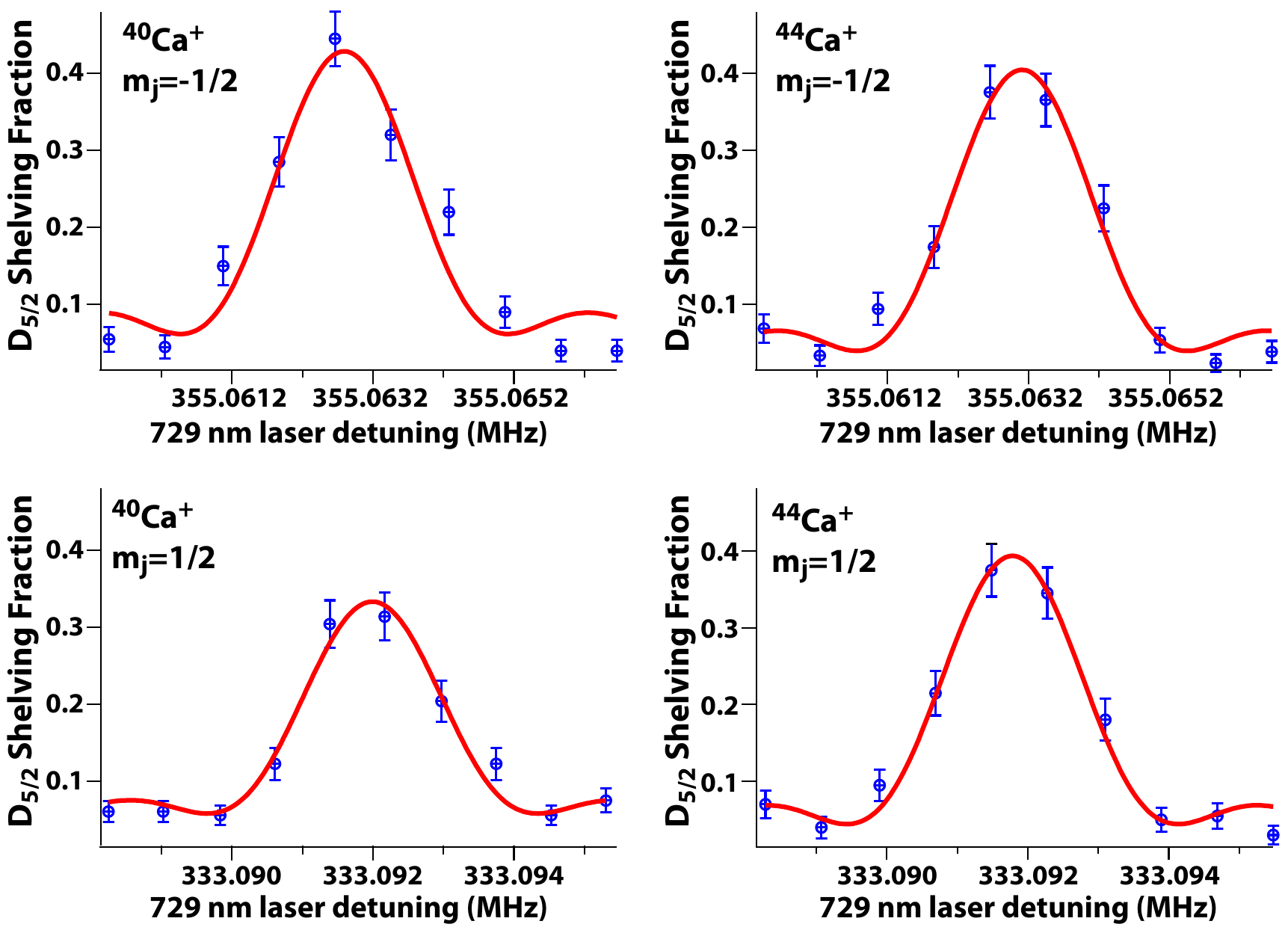}
\caption{\label{fig:spectra} (color online) A typical set of scans across the $\Delta m=0$ transitions showing simultaneous excitation of co-trapped $^{40}$Ca$^+$ and $^{44}$Ca$^+$, including data (blue) and fits (red).  Scans in each row are acquired simultaneously, with different EOM frequency sidebands addressing the $^{40,44}$Ca$^+$ ions such that they can both be excited at very nearly the same laser detuning.  The total time required to acquire the complete set of four scans is approximately thirty seconds.  We do not perform state preparation, thus ions are roughly equally likely to be found in the $|^2\textrm{S}_{1/2}, -\frac{1}{2}\rangle$ or $|^2\textrm{S}_{1/2}, \frac{1}{2}\rangle$ states prior to excitation.  As a consequence we expect a maximum $3^2$D$_{5/2}$ state shelving fraction of $\lesssim 50$\%, consistent with the data.  For the scans presented here the Zeeman shift between the two $\Delta m = 0$ transitions is roughly 0.9856~MHz, while $\pm$\nth{1} order EOM frequency sidebands at 2,670,443,750~Hz approximately span the transition isotope shift.}
\end{figure}

To execute a measurement of the $4^2$S$_{1/2} \rightarrow 3^2$D$_{5/2}$ transition isotope shift we first Doppler cool both isotopes for a period of roughly 1~ms.  All lasers addressing dipole-allowed transitions are then switched off, and the $4^2$S$_{1/2} \rightarrow 3^2$D$_{5/2}$ transition is probed using a pulse (typically 400~$\mu$s; laser-ion coherence is limited to $\sim$1~ms by magnetic field noise) from the 729~nm laser.  We set the drive frequency $\Delta f_{\mathrm{EOM}}$ for the 729~nm fiber EOM to very-nearly one-half the isotope shift while setting the EOM's modulation depth to maximize the power in the first-order frequency sidebands ($\beta \approx 1.8$; see Fig.~\ref{fig:apparatus} for an approximate modulation spectrum).  By tuning the laser frequency correctly using the $\sim$200~MHz AOM it is possible to scan the first-order EOM frequency sidebands across the $4^2$S$_{1/2} \rightarrow 3^2$D$_{5/2}$ transition in two isotopes simultaneously.  We probe the resulting 3$^2$D$_{5/2}$-state population as a function of 729~nm laser detuning for each isotope using electron shelving spectroscopy by making sequential 400~$\mu$s isotope-selective fluorescence measurements at 397~nm.  We typically detect $\sim$25 Poisson-distributed counts if an ion was not excited by the spectroscopy pulse and $<1$ count if the ion is excited,  yielding nearly perfect discrimination between `bright' and `dark' states.  After a 2~ms pulse from the 854~nm to empty population out of the $3^2$D$_{5/2}$ state, the entire sequence is repeated 200$\times$ at each detuning to build statistics,  yielding spectra like those shown in Fig.~\ref{fig:spectra}. 
 Fitting the resulting Fourier-limited spectra to $\mathrm{sinc}^2\bigl( (f_{\mathrm{AOM}}-f_0) \tau \bigr)$ functions to extract the peak-centers (where $\tau$ is the 729~nm laser pulse duration) yields the transition isotope shift: 
\begin{equation}
\Delta \nu_{\mathrm{IS}} = 2 \times \Delta f_{\mathrm{EOM}} + 2 \times \delta f_0.
\end{equation}  
Here $\delta f_0$ is the difference in fitted peak-centers determined from driving a particular $\Delta m = 0$ transition, while the factors of 2 reflect that we are using both the plus and minus first-order EOM sidebands and double-passing the 729~nm AOM, respectively.   However, in the case that the magnetic field is not precisely identical at the location of the two trapped ions, differential first-order Zeeman shifts are indistinguishable from the transition isotope shift.  To circumvent this problem we make near-simultaneous scans over two transitions with opposing magnetic field sensitivities, $|^2\textrm{S}_{1/2}, \frac{1}{2}\rangle \rightarrow |^2\textrm{D}_{5/2}, \frac{1}{2}\rangle$ and  $|^2\textrm{S}_{1/2}, -\frac{1}{2}\rangle \rightarrow |^2\textrm{D}_{5/2}, -\frac{1}{2}\rangle$ (Fig.~\ref{fig:transitions}), interlacing sequences of 729~nm laser pulses and detection between the two transitions on a roughly 5~ms timescale.  Fitting all four spectra, extracting isotope shifts for both transitions, and averaging the two results yields a measurement of the isotope shift which is first-order insensitive to spatial inhomogeneity in the magnetic field.  To remove sensitivity to linear drifts in experimental parameters we randomize the sequence of detunings used to scan across the transition; we also switch the order in which we probe the two transitions and detect  3$^2$D$_{5/2}$ state populations after every scan.  Each time an ion is lost from the trap we reload the two isotopes in the reverse order, alternating between ``AB'' and ``BA'' configurations of isotopes in the two 100~$\mu$m-separated minima of the axial double-well trapping potential.  This further mitigates possible systematic effects due to small environmental variations between the two ion locations.  
 
\section{Results}

\begin{figure}[ht!]
\includegraphics[width=8.0 cm]{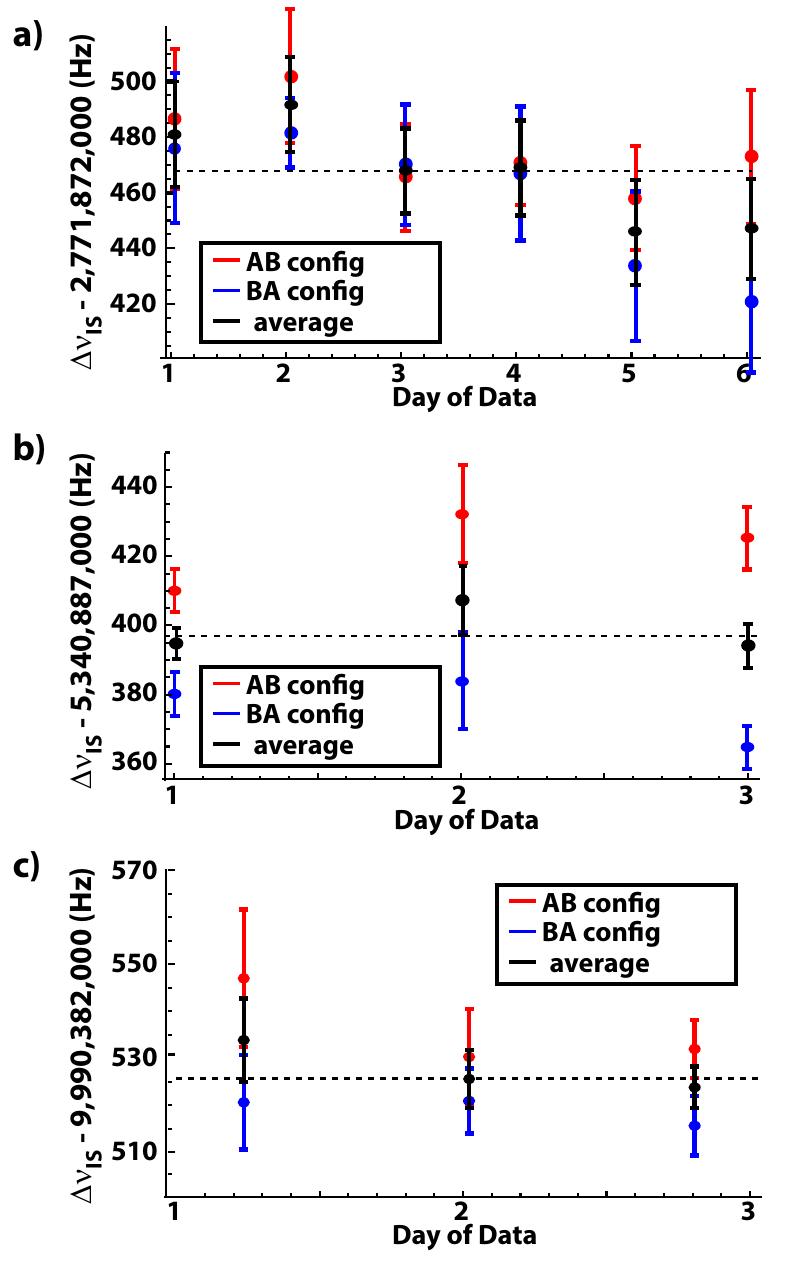}
\caption{\label{fig:results} (color online) Isotope shift data for (a) $^{40-42}$Ca$^+$,  (b) $^{40-44}$Ca$^+$,  (c) $^{40-48}$Ca$^+$.  Daily averages for the ``AB'' and ``BA'' ion configurations are shown in red and blue, respectively, while final combined determinations of the transition isotope shift are displayed in black.}
\end{figure}

Scans of the sort shown in Fig.~\ref{fig:spectra} yield a determination of the transition isotope shift with a statistical precision set by the quadrature sum of the uncertainty on the fitted peaks, typically 100-300~Hz.  As a consequence of collecting data with a given pair of ions until an ion is lost from the trap, the raw data is naturally organized into sequential blocks of measurements obtained in either the ``AB'' or ``BA'' ion configuration.  To reduce this collection of measurements into a final measured isotope shift we begin by excluding any scans for which the uncertainties in the fitted peak centers is larger than twice the median fit uncertainty, which typically indicates loss of an ion during the scan.  We then determine of a mean and standard deviation for each block.  A weighted average of the means of each day's blocks from a given configuration yields that day's value for the configuration (red and blue points in Fig.~\ref{fig:results}).  Daily values from the two configurations can then be combined in an unweighted average to obtain that day's final measured value for the transition isotope shift (black points in Fig.~\ref{fig:results}).  
Averaging between configurations for only one day's data at a time ensures that averages compare data for which the experimental conditions are as similar as possible.  Following this procedure yields isotope shifts of 2,771,872,467.6(7.1)~Hz,  5,340,887,396.6(3.5)~Hz, and 9,990,382,525.0(3.1)~Hz for  $^{40-42}$Ca$^+$,  $^{40-44}$Ca$^+$, and $^{40-48}$Ca$^+$, respectively.  These quoted errors, which are purely statistical, have been adjusted according to a $\chi^2$ analysis to account for observed scatter in the data.

\begin{figure}[ht!]
\includegraphics[width=7.7 cm]{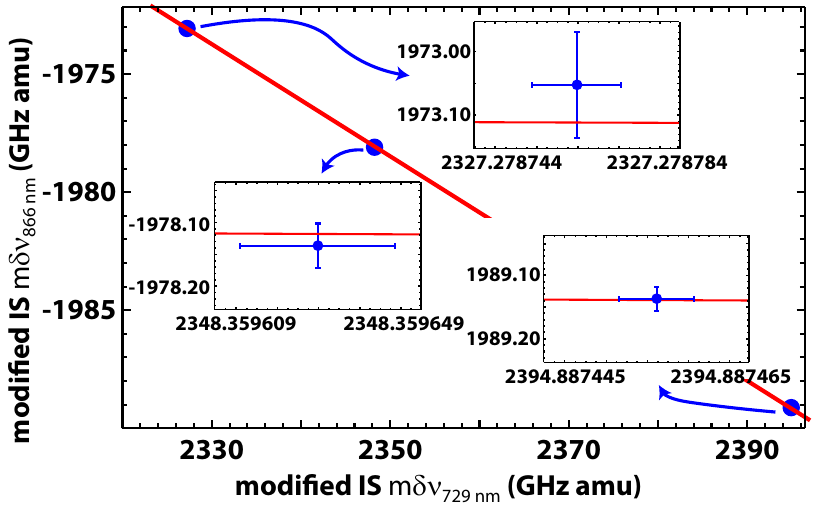}
\caption{\label{fig:KP} (color online) King Plot of the modified isotope shift $m\delta \nu_\lambda$ (see text).  Horizontal error bars include contributions from both the present measurement (see section~\ref{sec:Discussion}) and uncertainties in measured nuclear masses; note scales on vertical and horizontal axes.  866~nm transition data from~\cite{GWW15e}, and nuclear masses deduced from ~\cite{WAK17e, NIST}.  The best-fit line yields a slope of $-.23687(76)$ and a $y$-intercept of $-1.4218(18) \times 10^3$ GHz$\cdot$amu; $\chi^2=1.04$.}
\end{figure}

By combining these data with isotope shifts from a second transition, it is possible to build a linear relationship between the two transitions to form a 2D King plot (Fig.~\ref{fig:KP}).  A full discussion can be found in~\cite{GWW15e} and references therein.  In brief, we plot the modified isotope shift $m\delta \nu_\lambda^{A,A'} \equiv \delta \nu_\lambda^{A,A'} g^{A,A'}$ for two different transitions against one another, where $g = (m_A^{-1} - m_{A'}^{-1})^{-1}$ (notation from \cite{GWW15e}).  The plot fits well to a straight line~\cite{KA07}, confirming King's linearity at the level of accuracy dictated by the use of data from the dipole-allowed 866~nm transition for one the plot axes.  Notably, the precision of the modified isotope shifts on the horizontal axis (the $4^2$S$_{1/2} \rightarrow 3^2$D$_{5/2}$ transition measured here), are limited by mass uncertainties rather than spectroscopic precision~\cite{uncertainty}.   Future stringent tests of King's linearity in the calcium system will thus require both precise spectroscopic data from a second transition and also improved measurements of the calcium nuclear masses.

\section{Discussion of Systematic Errors}
\label{sec:Discussion}

Systematic errors affecting electric quadrupole transitions in single trapped ions have been carefully considered in the context of optical frequency standards; a particularly detailed treatment of the 674~nm electric quadrupole transition in $^{88}$Sr$^+$ is presented in \cite{MBD04e}.  This treatment may be mapped directly on to the 729~nm transition described here with the important caveat that, in the case of relative frequency measurements such as isotope shifts, many potential systematics are common-mode and thus cancel.  As a consequence, Hz-level precision may be achieved without requiring the same exquisite control that has been demonstrated in trapped-ion atomic clocks.  Here we present a thorough accounting of potential sources of error in the present isotope-shift measurement; a complete error budget is given below in Table~\ref{table:errors}.

\subsection{Stark Shifts}

Stark shifts, which arise due to differences in the polarizabilities of the $4^2$S$_{1/2}$ and $3^2$D$_{5/2}$ states, can appear in our experiment from three sources: DC Stark shifts due to the ions seeing non-zero time-averaged electric fields as a result of thermal- and micromotion, blackbody shifts due to the background room-temperature radiation; and AC Stark shifts due to off-resonant couplings to the 729~nm probe laser or ``leakage light'' from the 397, 866, and 854~nm lasers that address dipole-allowed E1 transitions.

\subsubsection{Thermal and Micromotion Stark shifts}
\label{sec:micro}

Trapped ions can experience a Stark shift due to their interactions with the electric field environment of the ion trap; secular and micromotion oscillations carry the ions away from the null of the RF electric field such that the time-averaged electric field they see is non-zero.  Secular motion at a temperature $T$ contributes a Stark shift for each ion of the form
\begin{equation}
(\Delta \nu_S)_{\mathrm{thermal}} = -\frac{\Delta \alpha_0}{2 h} \frac{3 M \Omega^2 k_B T}{e^2}.
\end{equation}
Here $\Omega$ is the trap's RF drive frequency, $\Delta \alpha_0$ is the difference in scalar polarizabilities between the $4^2$S$_{1/2}$ and $3^2$D$_{5/2}$ states \cite{MBD04e, ASC07} (taking the secular motion to be isotropic), and $M$ is the ion mass.  Each ion experiences a shift $(\Delta \nu_S)_{\mathrm{thermal}} \approx 10$~mHz for an ion temperature of $2\; T_{\mathrm{Doppler}}$.  However, most of this shift is common mode; for our measured temperature uncertainty of 50\% we find a differential shift of $2(1)\times 10^{-4}$~Hz per amu of mass difference between isotopes.  

In addition to their secular motion, the ions also sample a non-zero electric field as a result of micromotion driven by the trap's RF potential.  For the 
$\Delta m =0$ transitions driven here, 
\begin{multline}
(\Delta \nu_S)_{\mu} = \\ \frac{1}{h}\biggl(\frac{\Delta \alpha_0}{2} + \frac{\alpha_2}{5} (3 \cos^2 \theta -1) \biggr) \;\bigl\langle E(\mathbf{u},t)^2\bigr\rangle,
\end{multline}
where $\alpha_2$ is the tensor polarizability of the $3^2$D$_{5/2}$ state, $\langle E^2(\mathbf{u},t)\rangle$ is the average electric field amplitude at the ion position, and $\theta$ is the angle between the field and the quantization axis \cite{MBD04e, BMB98e}.  The electric field can be related to the ion velocity $V_\mu$ according to
\begin{equation}
\label{eqn:micromotion}
\bigl\langle E^2(\mathbf{u},t)\bigr\rangle = \biggl( \frac{M \Omega}{e}\biggr)^2 \bigl\langle V_\mu^2 \bigr\rangle,
\end{equation}  
thus it is possible to characterize the size and direction of the electric field the ion samples by measuring micromotion sidebands along three orthogonal directions \cite{BMB98e}.  

Stark shifts from ion micromotion provide the vast majority of the disagreement between isotope shift data collected in the ``AB'' and ``BA'' ion configurations (Fig.~\ref{fig:results}).  We typically observe differential Stark shifts of $\sim$10~Hz, as in Fig.~\ref{fig:results}(a,c) due primarily to a moderate amount of uncompensatable micromotion ($\beta \lesssim 0.4$) of unknown origin.  In this case the strength of the electric field sampled by the ion is independent of mass to leading order ~\cite{BMB98e}.  Differential Stark shifts are therefore canceled by averaging over the ``AB'' and ``BA'' ion configurations.  However, we observe that the amount of micromotion varies slowly with time.  Based on the typical time elapsed between measurements in each configuration and the rate at which we observe the micromotion environment to vary, we therefore include an uncertainty equal to $15\%$ of the characteristic difference between ``AB'' and ``BA'' configurations to account for imperfect cancelation when averaging.  

DC electric fields can also cause micromotion if they push the ion away from the trap's RF null.  Such fields can typically be well compensated, as they generally vary only over long timescales.  However, charging of trap surfaces arising from pointing instabilities of our 397~nm laser  led to larger than normal micromotion during some of our data collection, causing the larger differential Stark shifts evident in the $^{40-44}$Ca$^+$ data plotted in Fig.~\ref{fig:results}(b).  These shifts are well explained by spatially varying DC fields on the scale of $\sim$100-200~V/m oriented along the $\hat{y}$ direction (orthogonal to the trap plane, see Fig.~\ref{fig:apparatus}), a direction along which we monitor and re-compensate ion micromotion infrequently.  For this stray-field induced micromotion averaging over ``AB'' and ``BA'' ion configurations does \emph{not} entirely cancel the differential Stark shift, as in this case the shifts scale approximately as $M^2$ \cite{BMB98e}.  For the $^{40-44}$Ca$^+$ data we thus include a small systematic shift of 2~Hz in Table~\ref{table:errors} to account for the residual shift that remains after averaging.

\subsubsection{Blackbody Shift}

The Ca$^+$ blackbody shift has been calculated to be 0.38(1)~Hz at room temperature for $^{43}$Ca$^+$ \cite{ASC07}, and should be essentially independent of isotope.  Closely spaced ions will thus see blackbody shifts which are closely matched and also time invariant.  As such, after averaging the ``AB'' and ``BA'' ion configurations, the blackbody contribution to the Stark shift should be negligible.   

\subsubsection{AC Stark Shifts}

AC Stark shifts arise from perturbations due to the rapidly oscillating fields of optical radiation.  For far-off-resonant light detuned by a frequency $\delta$ these shifts are well described by
\begin{equation}
\label{eqn:AC_stark}
(\Delta \nu_S)_{AC} = \kappa \frac{I_0}{\delta},
\end{equation}
where $I_0$ is the laser intensity and $\kappa$ is a constant related to the transition's polarizability.  Such shifts can be measured explicitly by driving the $4^2$S$_{1/2} \rightarrow 3^2$D$_{5/2}$ transition in a single ion in the presence of (deliberately) leaked light.  

The 866~nm lasers are far (THz) detuned from any transitions involving the $4^2$S$_{1/2}$ or $3^2$D$_{5/2}$ states, thus the AC Stark shift they cause is too small to be measured even at full laser power.  As such, we bound rather than estimate the size of this shift.  For the 397 and 854 nm lasers, however, we measure AC Stark shifts as a function of laser power and detuning to confirm the scalings described by Eqn.~\ref{eqn:AC_stark}.  Extrapolating to the powers and detunings present during isotope shift measurements yields the values shown in table~\ref{table:errors}.

%  The 397~nm and 854~nm lasers caused shifts of 750~Hz and 11~kHz at 0.2$\mu$W and 300$\mu$W, respectively, at detunings characteristic of those present when conducting the $^{40-44}$Ca$^+$ isotope shift measurement.

\begin{figure}[ht!]
%May want to make this figure two-column
\includegraphics[width=8.5 cm]{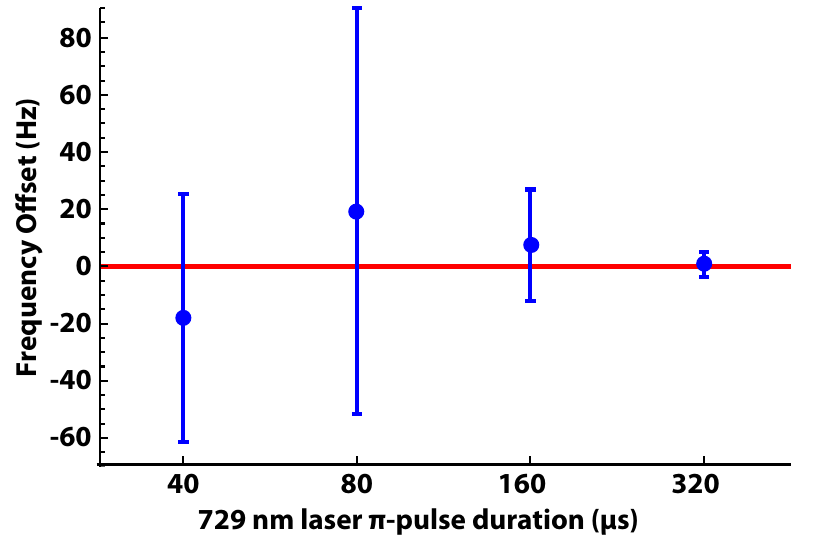}
\caption{\label{fig:AC_Stark} (color online) Variation in measured isotope shift for $^{40-44}$Ca$^+$ as a function of 729~nm laser $\pi$-pulse duration; 
$t_\pi \propto 1/\sqrt{\textrm{Power}}$.  The weighted average of the data is indicated by the red line.}
%  These data, were collected with the EOM frequency referenced to a clock with stability inferior to that of the Rb stabilized oscillator used for final data collection, and are not included in the analysis presented elsewhere in the article.}
\end{figure}

The 729~nm laser can also cause AC Stark shifts  \cite{HGR03e}.  We expect contributions due to off-resonant couplings to the numerous accessible Zeeman components (Fig.~\ref{fig:transitions}) of the $4^2$S$_{1/2} \rightarrow 3^2$D$_{5/2}$ transition and also as a result of the multiple off-resonant laser frequency components added by the fiber EOM.   Since the EOM sidebands are spaced by large (GHz) frequencies, Stark shifts due to the 729~nm laser are dominated by coupling of the resonant EOM sidebands to the nearby $\Delta m = \pm2$ transitions.  This principal piece of the shift, which can be calculated to be approximately 20~mHz for the 1.25~kHz Rabi frequency used in the experiment, is common to both isotopes and does not affect our results.  Further, any small differential contribution due to gradients in the laser intensity or Zeeman splittings between ions is cancelled when averaging over ion configurations.  To confirm this model, $^{40-44}$Ca$^+$ isotope shift data were collected as a function of 729~nm laser power (Fig.~\ref{fig:AC_Stark}), showing no resolvable shift.  However, we expect that small shifts of  $\lesssim 10^{-4}$~Hz from far-off-resonant EOM sidebands persist; we include these shifts in our final error budget in table~\ref{table:errors} along with corresponding uncertainties of 10\% of the shift size that stem from imperfect knowledge of the EOM modulation depth.

\subsection{Zeeman Shifts}

Probing the two co-trapped isotopes on a pair of transitions with symmetric magnetic field dependencies completely cancels DC linear Zeeman shifts, as any effects due to AC fields causing broadening and/or lineshape distortion should be common mode.  Differential quadratic Zeeman shifts between the two 100~$\mu$m spaced ions remain as a potential source of error, but these too should cancel when the ``AB'' and ``BA'' ion configurations are averaged together.  As such, we need only concern ourselves with differential quadratic Zeeman shifts which are also time-varying.  For the $\Delta m=0$ transitions probed here the quadratic Zeeman shift is given by \cite{MBD04e}
\begin{equation}
\Delta \nu_{Z2} = \frac{6}{25} \frac{\mu_B^2}{h^2\nu_{FS}} B^2
\end{equation}
where $\mu_B$ is the Bohr magneton, $B$ the ambient magnetic field, and $\nu_{FS}$ the fine-structure splitting between the $3^2$D$_{3/2,5/2}$ levels.  To consider a differential shift we write:
\begin{equation}
(B+\delta B)^2 - B^2 \approx 2 B \delta B + \ldots
\end{equation}
In our experiment the quantization field can be determined from the linear Zeeman shift of single ions, finding $B =  1.755(1)$~G; $\delta B = 1.4$~mG is the variation in the quantization field between the two 100~$\mu$m-spaced ions.  Conservatively assuming a 1\% time-variation in the quantization field yields an error of order $10^{-5}$~Hz.  Including magnetic contributions from room-temperature blackbody radiation leads to a negligible change in this result.

\subsection{Doppler Shifts}

Linear Doppler shifts are in general strongly suppressed for ions trapped in the Lamb-Dicke regime \cite{LME07}.  Small linear shifts could still occur if the trap fluctuates in a way that is correlated with the probe laser pulses~\cite{RHS08e}, but such shifts should be common to both ions and thus cancel; any differential shift is eliminated by averaging over ``AB'' and ``BA'' ion configurations.  However, \nth{2}-order Doppler shifts due to time dilation do result from both the ions' thermal secular motion and driven micromotion.  Thermal motion contributes a shift of
\begin{equation}
\Delta \nu_{\mathrm{thermal}} = -\nu_{729} \frac{3 k_B T}{2 M c^2},
\end{equation}
where $\nu_{729}$ is the $4^2$S$_{1/2} \rightarrow 3^2$D$_{5/2}$ transition frequency.  If the two trapped ions are at the same temperature this shift is largely common-mode; for $T = 2\;T_{\mathrm{Doppler}}$ the differential shift is approximately $3\times10^{-5}$~Hz/amu of mass difference between the ions.  However, since the ions may not be equally well-cooled, so based on our measured temperature uncertainty we conservatively assume a temperature difference of $1\times T_{\mathrm{Doppler}}$ and assign a corresponding uncertainty of 0.7~mHz.  

Driven micromotion contributes a shift of the form
\begin{equation}
\Delta \nu_{\mathrm{D2}} = - \frac{\nu_{729}}{2}\frac{\bigl\langle V_\mu^2 \bigr\rangle}{c^2}.
\end{equation} 
Estimates of $V_\mu$ determined from measurement of micromotion sidebands suggest \nth{2}-order Doppler shifts from micromotion of order 1~Hz.  However, these shifts are largely common-mode, and thus cancel in averages over ``AB'' and ``BA'' ion configurations; we include an uncertainty of 0.1~Hz to account for imperfect cancelation due to the slowly time-varying nature of the micromotion environment.

\begin{table*}[t]
	\begin{tabular}{ l || c | c || c | c || c | c }
		\begin{tabular}{p{2.4in}}
		\hfill \textbf{Isotope} \\ \hfill \textbf{Pairing}\\
		\end{tabular} & 
		\multicolumn{2}{c ||}{   \multirow{2}{*}{   \textbf{\large $^{40-42}$Ca$^+$}   }  } &
		\multicolumn{2}{c ||}{   \multirow{2}{*}{   \textbf{\large $^{40-44}$Ca$^+$}   }  } & 
		\multicolumn{2}{c}{       \multirow{2}{*}{   \textbf{\large $^{40-48}$Ca$^+$}   }  }\\
		
%		\multicolumn{2}{c ||}{\multirow{2}{*}{\textbf{\large $^{40-42}$Ca$^+$}}} & 
%			\multicolumn{2}{c ||}{\multirow{2}{*}{\textbf{\large $^{40-44}$Ca$^+$}}} &  
%			\multicolumn{2}{c}{\multirow{2}{*}{\textbf{\large $^{40-48}$Ca$^+$}}} \\
			
%		\textbf{pairing} &\multicolumn{2}{c ||}{ } & \multicolumn{2}{c ||}{ } &\multicolumn{2}{c}{ }\\
		\hline \hline
		\textbf{Source} & \textbf{Shift} &\textbf{Uncertainty} & \textbf{Shift} &\textbf{Uncertainty} & \textbf{Shift} &\textbf{Uncertainty}\\
		\hline

		%Blackbody contribution to Stark shift & 0 & $3.6\times10^{-3}$ & 0 & $3.6\times10^{-3}$ & 0 & $3.6\times10^{-3}$ \\
		Secular motion Stark shift & $4 \times 10^{-4}$ &  $2 \times 10^{-4}$  &  $9 \times 10^{-4}$ &  $4 \times 10^{-4}$  &  $1.7 \times 10^{-3}$ &  				$9 \times 10^{-4}$ \\
	
		Micromotion Stark shift & 0 & 2.5  & 2.0 & 6.8  &0 & 2.3 \\
		
		AC Stark shift due to light at & & & & & & \\
		
			\hspace{2em} 397~nm  & $2\times10^{-2}$ & $1\times10^{-2}$ & $2.7\times10^{-2}$ & $1.4\times10^{-2}$ & $1\times10^{-2}$ & 					$5\times10^{-3}$ \\
		
			\hspace{2em} 854~nm & $1.4\times10^{-2}$ & $3\times10^{-3}$ & $2.1\times10^{-2}$ & $4\times10^{-3}$ & $9.6\times10^{-2}$ & 					$1.9\times10^{-2}$ \\
			
			\hspace{2em} 866~nm &  $<1\times10^{-3}$ & $<1\times10^{-3}$ & $<1\times10^{-3}$ & $<1\times10^{-3}$ & $<1\times10^{-3}$ & 				$<1\times10^{-3}$ \\
			
			\hspace{2em} 729~nm & 3 $\times 10^{-4}$ & 1.5 $\times 						10^{-4}$ & 1.5$\times 10^{-5}$ & 8.2$\times 					10^{-5}$ & 8$\times 10^{-5}$ & 8$\times 10^{-6}$ \\

		\nth{2}-order Zeeman Shift & 0 & $10^{-5}$  & 0 & $10^{-5}$  & 0 & $10^{-5}$ \\
		
		\nth{2}-order Doppler & & & & & & \\
			\hspace{2em} Thermal motion&  $7\times10^{-5}$ & $7\times10^{-4}$ & $1.3\times10^{-4}$ & $7\times10^{-4}$ & $2.4\times10^{-4}$ & 				$7\times10^{-4}$ \\
			
			\hspace{2em} Micromotion  & 0 & 0.1  & 0 & 0.1  & 0 & 0.1 \\
		
		Electric quadrupole shift of $^2$D$_{5/2}$ level & 0 & 0.57 & 0 & 0.57 & 0 & 0.57 \\
		
		Rb reference aging & 0 & 0.83  & 0 & 1.60 & 0 & 3.00 \\
		Estimated Systematic Total &  3.4$\times 10^{-2}$ & 2.7  &  2.0 & 7.0 &  1.1$\times 10^{-1}$ & 3.8 \\
		\hline 
		Statistical & - & 7.1 & - & 3.5 & - & 3.1\\
		\hline \hline
		\textbf{Total} & $\mathbf{3.5\times 10^{-2}}$ & \textbf{7.6} & $\mathbf{2.0}$ & \textbf{7.8} & $\mathbf{1.1\times 10^{-1}}$ & \textbf{4.9}\\ 
	\end{tabular}
	\caption[Error Budget]{\label{table:errors}Estimated systematic frequency shifts and uncertainties (Hz).}			
\end{table*}

\subsection{Electric Quadrupole Shift}

The non-spherical nature of the $3^2$D$_{5/2}$ state gives rise to an electric quadrupole moment, measured experimentally to be $\Theta = 1.83(1) ea_0^2$ \cite{RCK06e}.  This moment can interact with electric field gradients to generate a frequency shift which is linear in the strength of the gradient,
\begin{equation}
\Delta \nu_{\mathrm{quad}} = \frac{\Theta}{h} \nabla E \biggl(\frac{1}{4} - \frac{3}{35} m_j^2 \biggr)(3 \cos^2 \theta -1),
\end{equation}
where $\theta$ is the angle between the quantization axis and the direction of the electric field gradient.  Rapidly oscillating gradients such as the trap's RF potential do not shift the line-center \cite{MBD04e}.  However, DC field gradients must be considered, including those responsible for axial confinement, breaking the degeneracy of radial secular modes, or arising from patch potentials present on the trap.  Differential shifts due to time-invariant electric field gradients, such as those from the axial trapping potential or the Coulomb field of the neighboring ion, cancel when averaging over ``AB'' and ``BA'' ion configurations.  But, time-varying patch potentials could cause non-zero shifts even after averaging.  We include a 0.57~Hz uncertainty, equivalent to the differential quadrupole shift which would be caused by a time-varying electric field gradient of $1~$V/m$\cdot \mu$m, twice that of our worst-case measured day-to-day stray-field variation.

\subsection{Rubidium Frequency Reference}

All RF frequencies in the experiment are derived from a Stanford Research SR625 10~MHz rubidium-referenced frequency standard.  The SR625 has a specified accuracy after factory calibration of $5\times10^{-11}$ and a long-term aging of $<5\times10^{-11}$/month.  All measurements reported here were completed within six months of factory calibration of the SR625, so we quote a fractional uncertainty of $3\times10^{-10}$ to account for aging of the rubidium reference.

\section{Conclusion}

We have completed a precise measurement of the $^{40-42,44,48}$Ca$^+$ isotope shifts in the $4^2$S$_{1/2} \rightarrow 3^2$D$_{5/2}$ electric quadrupole transition by simultaneously interrogating co-trapped ions using sidebands derived from a single laser source.  After incorporating all systematic shifts and uncertainties we find isotope shifts of 2,771,872,467.6(7.6), 5,340,887,394.6(7.8), and 9,990,382,525.0(4.9) Hz between $^{40}$Ca$^+$ and $^{42,44,48}$Ca$^+$, respectively.  Nearly all systematic shifts in this system are common-mode between the two trapped ions, thus improvements to laser-ion coherence and reduced ion micromotion should permit future reduction of uncertainty to below the 1~Hz level.  Similar tools and techniques can also be used to make an analogous measurement of isotope shifts in the $4^2$S$_{1/2} \rightarrow 3^2$D$_{3/2}$ transition, work we are now undertaking.  Our present measurement, when combined with existing or future isotope shift measurements as well as precise measurements of nuclear masses in the calcium system, can be used to search for King nonlinearities with unprecedented sensitivity.  Such searches are expected to set limits on new physics and/or provide a stringent test for calculations of King nonlinearities within the Standard Model.

\begin{acknowledgments}

Thanks to Cole Meisenhelder, Ariel Silbert, Sierra Jubin, and Sarah Stevenson, all of whom played essential roles in the assembly of the experiment, and to Tiku Majumder for helpful conversations.  This work was supported by the NSF under RUI grant number PHY-1707822 and by a Cottrell Scholar Award from the Research Corporation for Science Advancement.

\end{acknowledgments}

\bibliographystyle{apsrev}
\bibliography{reference_database}

\begin{thebibliography}{59}
\bibitem[{unc()}]{uncertainty}
\bibinfo{note}{Calcium atomic masses have been measured to a fractional
  precision of ($5.5\times10^{-10}$, $3.8\times10^{-9}$, $7.9\times10^{-9}$,
  $2.1\times10^{-9}$) for $^{40,42,44,48}$Ca, respectively. For the
  spectroscopic precision quoted here these mass uncertainties contribute
  between 82\% and 98\% of the total error budget for the modified isotope
  shifts plotted on the horizontal axis in Fig.~6.}
\end{thebibliography}

\end{document}